\documentclass[notitlepage, prd, nobalancelastpage, amssymb, superscriptaddress, showpacs, preprintnumbers, nofootinbib]{revtex4-2}

\usepackage{hyperref}
\hypersetup{colorlinks=true, citecolor=blue, urlcolor=blue, linkcolor=blue}

\usepackage{graphicx}
\usepackage{amsmath,amsthm,amssymb,bm,amsfonts}
\usepackage{todonotes}

\usepackage{color}
\usepackage[makeroom]{cancel}
\usepackage[normalem]{ulem}

\def\qbar{{\overline{q}}}

\def\fprime{{f^\prime}}

\def\Kcal{{\cal{K}}}

\def\eto{{\rm{e}}}

\def\bra{\big<}
\def\ket{\big>}

\def\qbar{{\bar{q}}}

\def\Tr{{\rm Tr\,}}

\newcommand{\bea}{\begin{eqnarray}}
\newcommand{\eea}{\end{eqnarray}}
\newcommand{\be}{\begin{equation}}
\newcommand{\ee}{\end{equation}}

\def\nsr{\langle n\rangle}

\begin{document}
\vspace*{1.5cm}
\title{\bf Partonic Entropy of the Proton from  DGLAP Evolution}

\author{Krzysztof Golec-Biernat}
\email{golec@ifj.edu.pl}
\affiliation{Institute of Nuclear Physics Polish Academy of Sciences, Radzikowskiego 152, 31-342 Cracow, Poland}

\begin{abstract}
We investigate the concept of partonic entropy of the proton within the Dokshitzer--Gribov--Lipatov--Altarelli--Parisi (DGLAP) evolution scheme of  collinear parton distributions.
We show that such entropy increases monotonically with the evolution scale. The mechanism underlying  this growth is illustrated 
using a simplified model of DGLAP evolution, highlighting the importance of including saturation effects at small $x$  in the evolution of parton distributions, which tame
the otherwise unbounded growth of partonic entropy.
Based on existing literature, we present two simplified models of parton saturation at small $x$. 
In one of these models, partonic entropy is identified with entanglement entropy and proposed as an experimentally  testable observable.
\end{abstract}


\maketitle
\section{Introduction}

The rapid growth of gluon distribution function for the Bjorken variable $x\to 0$, found in the deep inelastic scattering (DIS) experiments
at the HERA accelerator, stimulated enormous theoretical activity. This behavior had been anticipated by perturbative quantum chromodynamics (pQCD)  prior to the first data from HERA, through the all-order resummation of large logarithms $\log(1/x)$. These developments led to computational frameworks based on evolution equations such as the DGLAP \cite{Gribov:1972ri,Altarelli:1977zs,Dokshitzer:1977sg} or BFKL (Balitsky-Fadin-Kuraev-Lipatov)  equations \cite{Kuraev:1977fs,Balitsky:1978ic}. 
The resulting physical picture describes the nucleon as a dense system of partons (gluons and sea quark–antiquark pairs) probed in the small-$x$ regime of DIS.  A central feature of this regime is parton saturation, by which the rapid growth of parton densities at small-$x$ is moderated by gluon recombination effects. This phenomenon  induces non-linear corrections to  evolution equations that are linear  in parton distribution functions (PDFs). The effective QCD-based  theory of parton saturation, known as the Color Glass Condensate,  is presented in a comprehensive review   \cite{Gelis:2010nm}.

In recent years, there has been a growing interest in understanding the entropy of parton systems within the nucleon, as investigated through DIS  or hadron-hadron scattering 
\cite{Kutak:2011rb,Peschanski:2012cw,Stoffers:2012mn,Kharzeev:2017qzs,Hagiwara:2017uaz,Kovner:2018rbf,Armesto:2019mna,Tu:2019ouv,Duan:2020jkz,H1:2020zpd,Dvali:2021ooc,Liu:2022ohy,Liu:2022hto,Liu:2022qqf,Hentschinski:2022rsa,Hentschinski:2023izh,Hentschinski:2024gaa,Caputa:2024xkp,Ramos:2025tge,Kutak:2025syp,Hentschinski:2025pyq,Kutak:2025tsx,Grieninger:2025wxg,Kharzeev:2021nzh,Kharzeev:2026jkq}.  Of particular interest is the dense system of partons with small momentum fractions $x$, where saturation occurs. Several interesting proposals have been made in this context, the most intriguing of which is a conjecture relating the entropy of the produced hadrons to the entanglement entropy of the parton system probed in DIS at small $x$ \cite{Kharzeev:2017qzs}. This means that the ideas known from the studies of the foundation of quantum mechanics are successfully applied to the high-energy particle scattering \cite{Kharzeev:2021nzh}. Moreover, quantum information tools are employed to characterize dense partonic systems  \cite{Caputa:2024xkp,Kutak:2025syp}. 

However, most of the above exciting results were obtained within simplified models of dense gluonic systems, in which gluons are treated as 
zero-dimensional dipoles evolving with rapidity $y=\ln(1/x)\to \infty$. Thus, it is tempting to adopt a more systematic approach to the entropy of partonic systems, starting from the conventional collinear PDFs that evolve according to the DGLAP equations. 
From this perspective, parton saturation is a phenomenon that places an upper bound on the unlimited growth of parton entropy with the evolution scale in the DGLAP approach. This bound is intimately connected to the small-$x$ behavior of the PDFs, suggesting a deep link between parton dynamics and information-theoretic properties of hadronic matter.

The paper is organized as follows. In Section \ref{sec:partentr}, we define partonic entropy and prove that it is positive. 
In Section \ref{eq:dglap}, we present the DGLAP evolution equations in the form  most appropriate for proving that the partonic entropy is a monotonic function of the evolution parameter, which is done   in Section \ref{sec:proof}. In Section \ref{sec:relative}, we relate the partonic entropy to
the relative entropy known from the theory of information. In Section \ref{sec:interpret},  we present a simplified model of DGLAP evolution to illustrate that the saturation of the small-$x$ behavior of PDFs is essential for taming the unbounded growth of the partonic entropy with the evolution scale. 
For completeness, in Section \ref{eq:model1}, we present the dipole model developed in the literature to realize parton saturation, while in Section \ref{eq:model2}, we discuss a similar model in which the concept of entanglement entropy in high energy scattering  was introduced.

\section{Partonic entropy}
\label{sec:partentr}

For non-negative PDFs,  the sum of  the singlet quark distribution
\be\label{eq:singlet}
\Sigma(x,t) =\sum_i (q_i(x,t) + \qbar_i(x,t))
\ee
and the gluon distributions $g(x,t)$,  weighted by the parton longitudinal momentum fraction $x$, 
\be\label{eq:P}
P(x,t)= x\left(\Sigma(x,t)+ g(x,t)\right)
\ee
admits an interpretation  of a probability density on the interval $x\in[0,1]$. This follows from the facts that $P(x)$ is non-negative and satisfies the normalization condition 
 \be\label{eq:mom}
\int_0^1 dx P(x,t) =1
\ee 
The  factor $x$   in Eq.~(\ref{eq:P}) is essential:  without it neither the singlet nor  gluon distributions are normalizable over the interval $[0,1]$.
Here,  $i=1,\ldots,N_f$ denotes quark flavor and  $t=\ln(Q^2/Q_0^2)$ is the evolution variable, with  $Q^2$ being the resolution scale 
at which the proton is probed,  identified with the QCD factorization scale, and $Q_0^2$ an initial scale for the evolution.

 Relation (\ref{eq:mom}) is  known as  the momentum sum rule which states that partons carry the entire longitudinal momentum of the proton. In our approach, however,  this relation is interpreted as a normalization condition for the probability distribution $P(x,t)$.

Following the Gibbs-Shannon formula \cite{Shannon:1948dpw,Jaynes:1957zza} for entropy of a probability distribution  $\{P_n\}$, defined on a discrete set of states $n$\footnote{In quantum mechanics, $S$ is the von Neumann entropy, $S=-\Tr\rho\ln\rho$, computed in the basis that diagonalizes the  density operator $\rho$. },
\be\label{eq:gibbsentropy}
S=-\sum_n P_n\ln P_n\,\ge\, 0\,,~~~~~~~~~\sum_n P_n=1
\ee
the  partonic entropy of the proton  is given by
\be\label{eq:entropy}
S(t)=\int_0^1 dx\,P(x,t)\ln P(x,t)
\ee
In Eq.~(\ref{eq:gibbsentropy}), the minus sign is necessary to ensure positivity of $S$, since  $0<P_n\le 1$ and the logarithm  is  always negative. 
This is not the case for   the probability distribution $P(x,t)$,  which has no  the upper bound; consequently the logarithm in Eq.~(\ref{eq:entropy})  can be either positive or negative. In particular, when $P(x,t)<1$,  the entropy density, 
\be\label{eq:sdens}
s(x,t)\equiv P(x,t)\ln P(x,t)
\ee
is negative. However,  no minus sign is introduced on the right-hand side of Eq.~(\ref{eq:entropy}).  because $S(t)$  is always  non-negative. 

This can be readily proven using a well-known inequality
\be\label{eq:ineq}
x\ln\!\left(\frac{x}{y}\right)\ge (x-y)\,,~~~~~~x\ge 0,\,y>0
\ee
 with the substitution: $x\to P(x,t)$ and $y\to 1$. Thus
\be
P(x,t)\ln P(x,t) \ge (P(x,t)-1)
\ee
and due to normalization condition (\ref{eq:mom}), one obtains
\be
S(t) \ge  \int_0^1dx\,(P(x,t)-1)=0
\ee
Note that the non-negativity is also a consequence of  the integration range  $x\in  [0,1]$. For different integration domains, the generally defined entropy (\ref{eq:entropy}) may admit a negative lower bound. 
This is in contrast to relative entropy, to be discussed later, which is always non-negative.

In the forthcoming analysis, it will be shown  that under the DGLAP  evolution of the PDFs \cite{Gribov:1972ri,Altarelli:1977zs,Dokshitzer:1977sg},  the partonic entropy 
$S(t)$ increases monotonically with  $t$, i.e.,  
\be\label{eq:posititvity}
\frac{dS}{dt} >0
\ee
For this purpose,  entropy (\ref{eq:entropy}) is differentiated with respect to $t$:
\be
\frac{dS}{dt} = \int_0^1dx\,\frac{\partial P(x,t)}{\partial t}\left(\ln P(x,t)+1\right)
\ee
The integration over one yields zero, since from Eq.~(\ref{eq:mom}) follows:
\be
\int_0^1 dx\,\frac{\partial P(x,t)}{\partial t}= \frac{d}{dt}\int_0^1 dx\,P(x,t)=0
\ee
and finally
\be\label{eq:pochodna}
\frac{dS}{dt} = \int_0^1dx\,\frac{\partial P(x,t)}{\partial t}\,\ln P(x,t)
\ee
To proceed, we  need to consider  the DGLAP equations  governing  the evolution of $P(x,t)$ with respect to $t$.

\section{DGLAP evolution equations}
\label{eq:dglap}

The general form of the DGLAP  evolution equations is given by  \cite{Golec-Biernat:2006qye,Golec-Biernat:2007rcl}:
\be\label{eq:dglap1}
\partial_t D_f(x,t)=\sum_{\fprime}\int_0^1 du\,\Kcal_{f\fprime}(x,u,t) D_\fprime(u,t)
\ee
Here,  $\partial_t=\partial/\partial t$ while  $D_f\in\{q_i,\qbar_i,g\}$ denotes the parton distribution functions 
(not multiplied by $x$). The kernel
\be\label{eq:kernel1}
\Kcal_{f\fprime}(x,u,t) =K^R_{f\fprime}(x,u,t)-\delta(x-u)\,\delta_{f\fprime}K^V_\fprime(u,t)
\ee
contains the real $R$ and virtual $V$ contribution to parton emission. The real part is computed in perturbative QCD  and corresponds to the transition of a parton with longitudinal momentum fraction and flavor $(u,\fprime)$  to a parton $(x,f)$. The virtual part leaves the parton $(u,\fprime)$ unchanged and is determined from the momentum sum rule:
\be\label{eq:sum1}
\frac{d}{dt} \int_0^1dx\,\sum_f xD_f(x,t)=0
\ee
which is valid if
\be
 \int_0^1dx\,\sum_f x\Kcal_{f\fprime}(x,u,t)=0
 \ee
Substituting (\ref{eq:kernel1}) into the above equation, one obtains:
\be
uK^V_\fprime(u,t) =  \int_0^1dx\,\sum_f x K^R_{f\fprime}(x,u,t)
\ee
With this result, the evolution equation (\ref{eq:dglap1}) can be written the following form
\begin{align}\nonumber
\label{eq:dglap2}
\partial_t D_f(x,t) &=\sum_{\fprime}\int_0^1 du\,K^R_{f\fprime}(x,u,t)D_\fprime(u,t)
\\
&-D_f(x,t)\sum_\fprime\int _0^1 du\left(\frac{u}{x}\right)K^R_{\fprime f}(u,x,t)
\end{align}
The real-emission kernel is given by
\be\label{eq:apkernels}
K^R_{f\fprime}(x,u,t) = \frac{1}{u}P_{f\fprime}(x/u,t)\,\theta(u-x)
\ee
 where  $P_{f\fprime}$ are the Altarelli--Parisi splitting functions, computed perturbateively as a series in the strong coupling constant
 $\alpha_s$:
\be
P_{f\fprime}(z,t) =\frac{\alpha_s(t)}{2\pi}P_{f\fprime}^{(0)}(z) + \frac{\alpha_s^2(t)}{4\pi^2} P_{f\fprime}^{(1)}(z)+\ldots
\ee
Substituting (\ref{eq:apkernels}) into (\ref{eq:dglap2}), one obtains:
\begin{align}\nonumber
\partial_t D_f(x,t) &=\sum_{\fprime}\int_x^1 du\,P_{f\fprime}(x/u,t)D_\fprime(u,t)
\\
&-D_f(x,t)\sum_\fprime\int _0^x \frac{du}{x}\left(\frac{u}{x}\right)P_{\fprime f}(u/x,t)
\end{align}
Changing the variables to $y=x/u$ in the first integral and to $y=u/x$ in the second one, one finds:
\begin{align}\nonumber
\partial_t D_f(x,t) &=\sum_{\fprime}\int_x^1 \frac{dy}{y}\,P_{f\fprime}(y,t)D_\fprime(x/y,t)
\\
&-D_f(x,t)\sum_\fprime\int _0^1 dy\,yP_{\fprime f}(y,t)
\end{align}
which can finally  be written as follows:
\begin{align}\nonumber
\label{eq:dglap3}
\partial_t D_f(x,t) &=\sum_{\fprime}\int_0^1\!dz\int_0^1\!dy\,\delta(x-yz)\,P_{f\fprime}(y,t)D_\fprime(z,t)
\\
&-D_f(x,t)\sum_\fprime\int _0^1 dy\,yP_{\fprime f}(y,t)
\end{align}
These are the DGLAP evolution equations. It is easy to show that they satisfy the momentum sum  relation (\ref{eq:sum1}).

\section{Proof of inequality (\ref{eq:posititvity})}
\label{sec:proof}

In the notation introduced in the previous section, the distribution $P(x,t)$ is given by the following sum
\be\label{eq:26abc}
P(x,t)=\sum_f xD_f(x,t)
\ee
where the summation includes quark and antiquark flavor and gluon. 
From Eq.~(\ref{eq:dglap3}),  one obtains
\begin{align}\nonumber
\label{eq:dglap4}
\frac{\partial P(x,t)}{\partial t} &=\sum_{f,\fprime}\int_0^1\!dz\int_0^1 \!dy\/x\,\delta(x-yz)\,P_{f\fprime}(y,t)D_\fprime(z,t)
\\
&-\sum_f xD_f(x,t)\sum_\fprime\int _0^1 dy\,yP_{\fprime f}(y,t)
\end{align}
Substituting this result in Eq.~(\ref{eq:pochodna}) and performing  the integration with the delta function, one finds:
\begin{align}\nonumber
\frac{dS}{dt} &= \sum_{f,\fprime}\int_0^1dz\int_0^1 dy\,yP_{f\fprime}(y,t)\,zD_\fprime(z,t)\ln P(yz,t)
\\
&-\sum_{f,\fprime} \int_0^1dx \int_0^1dy\,yP_{\fprime f}(y,t)\,xD_f(x,t)\,\ln P(x,t)
\end{align}
Renaming the summation indices $f\leftrightarrow\fprime$ and changing the integration variable $x\to z$ in the second term on the r.h.s., 
one   finally finds\footnote{Note that the pole singularity at $y=1$  in the diagonal splitting functions $P_{ff}(y)$ is regularized  by the logarithmic function, in accordance with the plus prescription for $y$ variable.}: 
\begin{align}\nonumber
\label{eq:maineq}
\frac{dS}{dt} &= \sum_{\fprime} \int_0^1 dz\,zD_\fprime(z,t) 
\\
&\times \int_0^1dy\, \Big(\sum_f yP_{f\fprime}(y,t)\Big)\,\ln\!\left(\frac{P(yz,t)}{P(z,t)}\right)
\end{align}
This is the main formula of the proof, which will be further analyzed.

\begin{figure}[t]
  \centering
 \includegraphics[width=9cm]{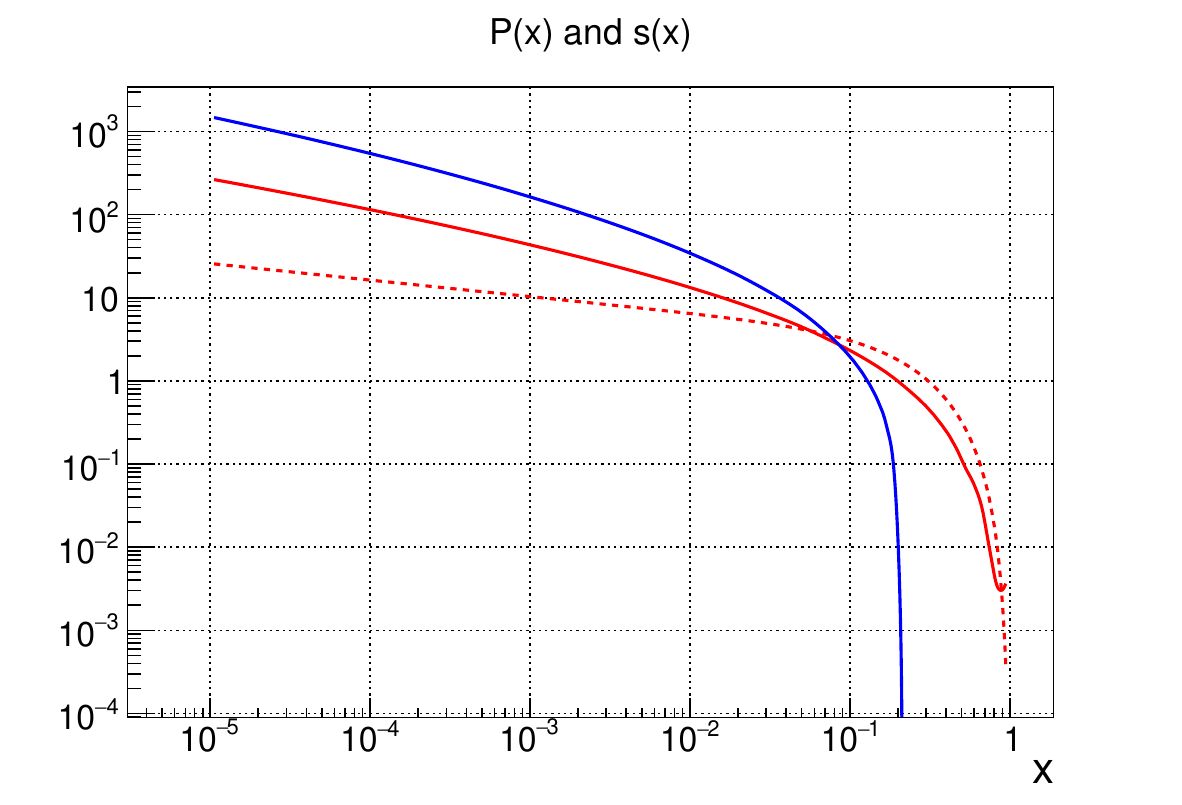}
   \caption{$P(x,t)$ given by Eq.~(\ref{eq:P}) as a function of $x$ at $Q^2=1,100~{\rm GeV}^2$ (dashed and solid red lines, respectively). The blue line shows the entropy density $s(x,t)$ given by Eq.~(\ref{eq:sdens}) at $Q^2=100~{\rm GeV}^2$. The sharp drop near $x\approx 0.2$ indicates that $s(x,t)$ becomes negative  when $P(x,t)<1$. 
 }
 \label{fig:1}
 \end{figure}

For non-negative parton distributions $D_f$ and splitting functions $P_{ff^\prime}$, only the logarithmic function on the r.h.s. of  Eq.~(\ref{eq:maineq}) needs to be examined.
From  global  DGLAP-based fits to hard scattering data, the singlet quark and gluon distributions present in the distribution $P(x,t)$, see Eq.~(\ref{eq:P}), are monotonically  decreasing functions of  $x\in [0,1]$, for the values of the evolution parameter $t=\ln(Q^2/Q_0^2)$ corresponding to  $Q^2$  sufficiently larger than the initial scale  $Q_0^2$, where artifacts of an  initial parameterization of PDFs  may spoil this monotonic behavior. 
As a consequence, we find
\be
P(yz,t)> P(z,t)
\ee
for $yz\le z$ for $y\in(0,1)$ (see the red lines in Fig.~1 where $P(x)$ is shown for $Q^2=1,100~{\rm GeV}^2$).   Therefore  the logarithmic factor in Eq.~(\ref{eq:maineq}) is strictly positive, 
\be
\ln\left(\frac{P(yz,t)}{P(z,t)}\right)> 0
\ee
Taking this relation into account in Eq.~(\ref{eq:maineq}), we find that
\be\label{eq:31rise}
\frac{dS}{dt}>0
\ee
Thus,  the partonic entropy increases monotonically under DGLAP evolution. 

\begin{figure}[t]
  \centering
  \includegraphics[width=9cm]{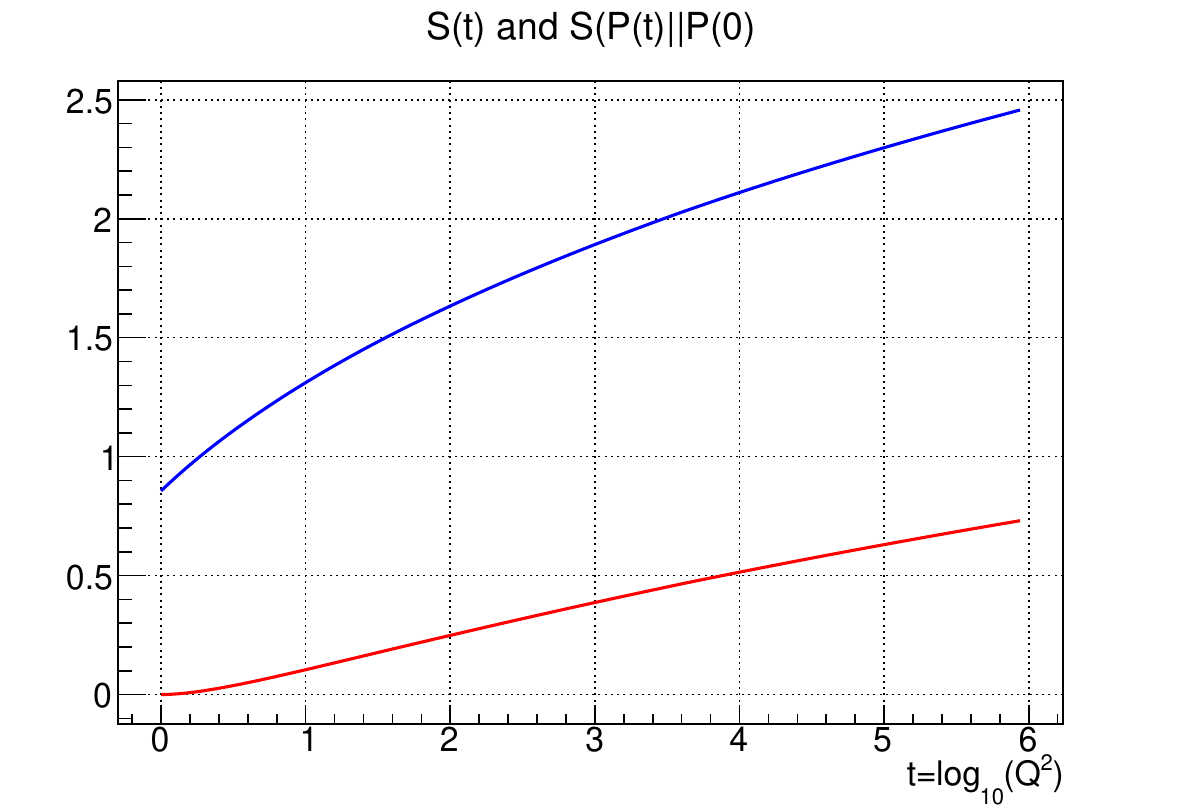}
   \caption{Partonic entropy (\ref{eq:entropy})  (upper blue line) 
 and relative entropy (\ref{eq:pentropy}) (lower red line) 
 as a function of  the evolution parameter $t=\log_{10}(Q^2/1\,{\rm GeV}^2$).
 }
 \label{fig:2}
 \end{figure}

This behavior is illustrated by   the upper blue line in Fig.~\ref{fig:2}, which corresponds to the entropy
\be
\label{eq:Sbiggger}
S_>(x_{\rm min})\equiv\int_{x_{\rm min}}^1 dx\,P(x)\ln P(x)
\ee
with $x_{\rm min}=10^{-8}$, due to numerical limitations. The missing contribution from the range $[0,x_{\rm min}]$ leads to an even stronger
growth with $t$,  and  is  mainly responsible for the unbounded increase of the partonic entropy,  see Section \ref{sec:interpret} for more detailed discussion of this result.

\section{Relative entropy}
\label{sec:relative}

It is insightful to examine the increase of the  partonic entropy  in terms of the relative entropy,  which for the discrete probability distributions
 $p_n$ and $q_n$ is defined as follows:
 \be\label{eq:KL}
S(p\|q) = \sum_n p_n\ln\!\left(\frac{p_n}{q_n}\right)
\ee
The relative entropy is non-negative\footnote{The quantum relative entropy,  $S(\rho\|\sigma)=\Tr\rho(\ln\rho-\ln \sigma)$ for any density operators $\rho,\sigma$, is also non-negative,
 see  e.g. \cite{Witten:2018zva} for the proof. }:
\be
S(p\|q) \ge 0\,,
\ee
which can be easily proven  using inequality  (\ref{eq:ineq}):
\be
\sum_n p_n\ln\!\left(\frac{p_n}{q_n}\right) \ge \sum_n (p_n-q_n)=0
\ee
where we used the normalization of the probability  distributions to unity.
In addition, $S(p\|q)=0$ if and only if  $p_n=q_n$ for all $n$.  Thus,  it is tempting to interpret $S(p\|q)$ as a metric. However, the relative entropy  is neither symmetric, $S(p\|q)\ne S(q\|p)$,  nor  does it satisfy the triangle inequality, 
therefore, it should  be  understood as a measure of distinguishability of the probability distribution $p_n$ from a reference distribution $q_n$,    rather than a distance between them.

For continuous probability distributions $P(x)$ and $Q(x)$, defined on the interval  $x\in[a,b]$ (finite or infinite), the relative entropy is given by
\be\label{eq:relent}
S(P\|Q)=\int_a^b dx\,P(x)\ln\!\left(\frac{P(x)}{Q(x)}\right) 
\ee
Comparing (\ref{eq:relent}) with (\ref{eq:entropy}), we see that the partonic entropy $S(t)$ can be interpreted as the relative
entropy of the parton distribution function $ P(x,t)$ with respect to the  uniform distribution $Q(x)\equiv 1$ on the interval $[0,1]$:
\be
S(t)=S(P(t)\|1)
\ee
Therefore, the unbounded growth of $S(t)$ in the DGLAP evolution scheme  characterizes   the increasing distinguishability  of 
the parton distribution $P(x,t)$  from the uniform distribution as $t$ increases.

We can also define  the relative entropy of  $P(x,t)$ with respect to the initial parton distribution at  $t=0$, $P(x,0)$:
\be\label{eq:pentropy}
S(P(t) \| P(0)) = \int_0^1 dx\,P(x,t) \ln\!\left(\frac{P(x,t)}{P(x,0)}\right)
\ee
which  quantifies  the distinguishability of the final parton distribution from the initial one, see the lower red line in Fig.~\ref{fig:2}. 
Such quantity, called dynamical entropy,  was introduced in \cite{Peschanski:2012cw} for different parton distributions.

In the hypothetical situation (which does not occur for DGLAP evolution) where the distribution $P(x,t)$ tends to an asymptotic form $P^{\rm as}(x)$ as $t\to \infty$, the relative entropy of $P$ with respect to $P^{\rm as}$ vanishes in this limit:
\be\label{eq:limit}
\lim_{t\to\infty} S(P(t))\|P^{\rm as})=0
\ee
Such behavior occurs, for example, for the pion distribution amplitude, $\phi(x,t)\equiv \sqrt{P(x,t)}$, which undergoes  the  Efremov--Radyushkin--Brodsky--Lepage (ERBL) evolution  \cite{Efremov:1979qk,Lepage:1980fj}, with the asymptotic form  $\phi^{\rm as}(x)=6\/x(1-x)$.

\section{Entropy growth interpretation}
\label{sec:interpret}

The unbounded increase  of partonic entropy originates from the growing number of partons resolved  by the electroweak boson  in deep inelastic scattering (DIS)
as the resolution scale, $\lambda=1/\sqrt{Q^2}\to 0$, approaches zero. These partons populate an expanding phase space for the emission of soft (wee) gluons carrying a small momentum fraction $x$.   Consequently, the sea-quark and gluon distributions rise rapidly   as $x\to 0$ and $Q^2\to \infty$.

This effect  can be qualitatively  illustrated by a simplified form of the solution to the DGLAP equations:
\be
P(x,t)= A(t)\, x^{\alpha(t)-1}(1-x)^{\beta(t)-1}
\ee
where the amplitude $A(t)$ is fixed by the normalization condition (\ref{eq:mom}), and  is given by the inverse of the  beta function
\be\label{eq:Anorm}
A(t)=\frac{1}{B(\alpha,\beta)}=\frac{\Gamma(\alpha(t)+\beta(t))}{\Gamma(\alpha(t))\,\Gamma(\beta(t))}
\ee
The convergence of the normalization integral requires $\alpha(t)>0$, while  the properties of the DGLAP equations imply $\beta(t)>1$.  
In the presented example, DGLAP evolution is simulated  by the limit $\alpha(t)~\to~0^+$ as $t\to \infty$, which 
results in a  strongly increasing parton distribution as $x\to 0$ and $Q^2\to \infty$.

\begin{figure}[t]
  \centering
 \includegraphics[width=9cm]{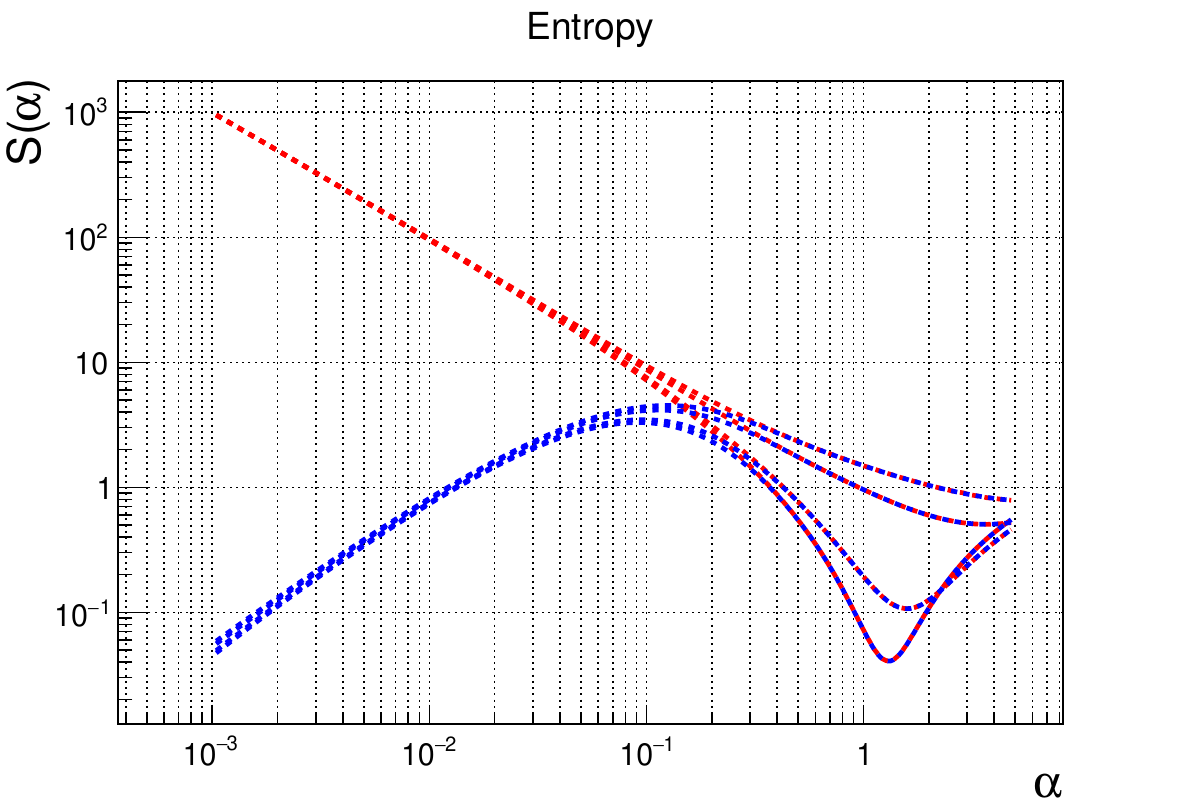}
 \caption{Red dotted lines: entropy (\ref{eq:Sab}) as a function of $\alpha$ for $\beta-1=0.2,1,5,10$ (from bottom to top).  
Blue dotted lines:  entropy (\ref{eq:Sbiggger}) for $x_{min} = 10^{-8}$ (this choice is for  illustration only) and  the above values of $\beta$. 
}
 \label{fig:3}
 \end{figure}

The partonic entropy (\ref{eq:entropy}) then takes the following  form 
\begin{align}\nonumber
S = \ln A &+(\alpha-1)\,\psi(\alpha)+(\beta-1)\,\psi(\beta)
\\\label{eq:Sab}
& - (\alpha+\beta-2) \,\psi(\alpha+\beta)
\end{align}
where $\psi(z)=\Gamma^\prime(z)/\Gamma(z)$ is the digamma function. Using the property of $\psi$:
\be
\psi(\alpha+1) = \psi(\alpha)+\frac{1}{\alpha}
\ee
one finds the following singularity structure of $S$  in the vicinity of  $\alpha\approx 0$:
\be
S= \ln\alpha+\frac{1}{\alpha} -1+\gamma+\psi(\beta)+{\cal O}(\alpha)
\ee
where $\gamma$ is the Euler's constant.
Thus,  as $\alpha \to 0^+$, the entropy increases without bound:
\be\label{eq:37}
S\simeq  \ln\alpha+ \frac{1}{\alpha} \simeq \frac{1}{\alpha}\to \infty
\ee
This behavior is  illustrated in Fig.~\ref{fig:3} by the red dotted lines, showing  $S(\alpha)$   for four values of $\beta$.

It can be demonstrated analytically that the singular behavior (\ref{eq:37})  results from the small-$x$ contribution to $S$:
\be
S_<(x_{\rm min})\equiv \int_0^{x_{\rm min}} dx\,P(x)\ln P(x) 
\ee
Indeed,  using  $P(x)= A\, x^{\alpha-1}$  for $x\le x_{\rm min}\ll 1$,  one obtains in the limit $\alpha\to 0^+$;
\be\label{eq:40}
S_<(x_{\rm min}))\simeq\ln\alpha+\frac{1}{\alpha}
\ee
This is illustrated by the blue dotted lines in Fig.~\ref{fig:3},  corresponding to the entropy $S$ without the small-$x$ contribution,
defined by Eq.~(\ref{eq:Sbiggger}). It clearly does not show the singular behavior.
Thus,  the small-$x$ partons  generate  an unbounded entropy.

This  observation raises a question about  saturation effects   which involve not only gluon splitting but also
gluon recombination \cite{Gribov:1983ivg,Mueller:1985wy,Mueller:1993rr,Mueller:1994gb,Mueller:1999wm,Kovchegov:1999yj,Kovchegov:1999ua}.   The latter phenomenon 
suppresses the small-$x$ increase of  the parton distributions  and hopefully  imposes an upper bound on  the  partonic entropy, see \cite{Kutak:2011rb} and the recent discussion in \cite{Dvali_2022} in the context of the Color Glass Condensate (CGC) \cite{Gelis:2010nm}, an effective theory of gluon saturation.

In our model, the PDFs suppression can be accomplished by freezing  the parton distribution $P(x)$  for $x<x_{\rm min}\ll 1$,
\be\label{eq:40new}
P(x) \equiv  (A/\alpha)\,x_{\rm min}^{\alpha-1}
\ee
where the additional factor of $1/\alpha$  is introduced to keep the small-$x$ contribution  to the momentum sum rule  unchanged, 
\be\label{eq:momx}
\int_0^{x_{\rm min}} dx\,P(x)  = (A/\alpha)\, x_{\rm min}^\alpha 
\ee
The  corresponding contribution to the entropy  is
\be
S_<(x_{\rm min}) = x_{\rm min} P(x_{\rm min})\ln P(x_{\rm min})
\ee
 and for  $\alpha\to 0^+$, one obtains a non-singular contribution 
\be
S_<(x_{\rm min}) \simeq  \ln (1/x_{\rm min}) + {\cal O}(\alpha)
\ee
Therefore, the  partonic entropy $S$ is no longer singular in the considered limit and has an upper bound which saturates its value.  Notice hat  the small-$x$
contribution (\ref{eq:momx}) also saturates the momentum sum rule since in the limit $\alpha\to 0^+$
\be
 (A/\alpha)\, x_{\rm min}^\alpha \simeq      x_{\rm min}^\alpha\to 1
\ee

A more refined approach to parton saturation should take into account the presence of a saturation scale $Q=Q_s(x)$, which in our example leads to a dependence of  $x_{\rm min}$ on the evolution parameter $t$.   This important aspect of parton saturation, however,  needs a more detailed analysis based on  non-linear QCD evolution equations, which lies beyond the scope of the present work.

\section{Reaction-diffusion model of parton saturation}
\label{eq:model1}

For the completeness of the discussion, we will present  simplified models of parton saturation effects at small $x$, discussed in the literature, which illustrate the saturation of entropy  as $x\to 0$. Strictly speaking, this is a  discussion of the Regge limit of QCD, in which the hard scale $Q^2$ is kept fixed while the invariant energy of the partonic processes $\sqrt{s}\to \infty$. Thus, the relevant quantities are parton distributions that depend not only on parton's momentum fraction $x$ but also on its transverse momentum $k_\perp$. However, in the simplified models presented below, the latter dependence is neglected.

In the reaction-diffusion  model, partons are gluon dipoles that are  zero-dimensional in the transverse direction. They evolve in the longitudinal direction through the dependence of  parton cascades on rapidity $y=\ln(1/x)$  \cite{Mueller:1994gb},  The evolution  incorporates  both parton splitting and recombination processes which  correspond to pomeron loops within the effective reggeon field-theory description, see \cite{Bondarenko:2006rh} for more details and further references.  
 
 The underlying dynamics
is represented as a reaction-diffusion Markov process with  the following master equation for the probability $p_n(y)$ of finding  $n$ partons:
\begin{align}\nonumber
\label{eq:sateq}
\frac{dp_n}{dy} &= \alpha (n-1)p_{n-1} -\alpha np_n 
\\
&+\beta n(n+1)p_{n+1}-\beta n(n-1) p_n
\end{align}
The first line on the right-hand side  describes   parton splitting $(n-1)\to n$ with rate $\alpha$ while the second line corresponds to parton recombination $(n+1)\to n$ 
with rate $\beta$.  
The two rates are related by  $\beta=\alpha_s^2\,\alpha$, where $\alpha_s\ll 1$. 
Enforcing the  condition $p_{n<0}=0$ implies that $p_0={\rm const}$, which is  then fixed by the  normalization condition:
\be
\sum_{n=0}^\infty p_n(y)=1
\ee
It is straightforward to check that Eq,~(\ref{eq:sateq}) admits a stationary solution for which $dp_n/dy=0$,  given by the Poisson distribution:
\be\label{eq:poisson}
P_n= \frac{N^n}{n!}\,e^{-N}
\ee
with the mean multiplicity equal to $N$,
\be\label{eq:meanN}
\bra n\ket\equiv \sum_{n=0}^\infty n\/P_n = N=\frac{\alpha}{\beta}=\frac{1}{\alpha_s^2}\gg 1
\ee

An approximate solution to Eq.~(\ref{eq:sateq}) for large $y$ was found in  \cite{Hagiwara:2017uaz} via  Borel summation:
\be
p_n(y) =  \frac{N^n}{n!}\,\int_0^\infty dz\,\eto^{-z-\frac{NzX}{1+zX}}\,\left(\frac{zX}{1+zX}\right)^n
\ee
with $X=\alpha_s^2\, y$ and the initial condition $p_n(0)=\delta_{n0}$. This distribution approaches  the Poisson distribution $P_n$
 as $y\to \infty$, 
 so that the relative entropy of $p_n(y)$ with respect to  $P_n$ vanishes in this limit:
\be
\lim_{y\to \infty}S(p(y)\|P)=0
\ee

Likewise,  as  demonstrated numerically in \cite{Hagiwara:2017uaz}, the partonic entropy
\be\label{eq:sy}
S(y) =-\sum_{n=0}^\infty p_n(y)\ln p_n(y)
\ee
approaches {\it non-monotonically} the entropy  of the Poisson distribution in the limit $y\to \infty$,  
\be
S(y) \to S_P\approx  \frac{1}{2}\,\ln(2\pi e N)-\frac{1}{12N} +{\mathcal O}(1/N^2)
\ee
while the mean multiplicity approaches    {\it monotonically}  the Poisson  distribution mean multiplicity:
\be
\bra n\ket(y) =\sum_{n=0}^\infty n p_n(y) \to N
\ee
Thus, both the partonic entropy and the mean mul\-ti\-pli\-ci\-ty saturate.

\section{Entanglement entropy in DIS} 
\label{eq:model2}

The reaction-diffusion model of parton saturation should be contrasted with the Markov process that includes only parton splittings,
see e.g. \cite{Mueller:1994gb,Levin:2003nc}, given by the first line of Eq.~(\ref{eq:sateq}),
\be\label{eq:bfklmodel}
\frac{dp_n}{dy} = \alpha (n-1)p_{n-1} -\alpha np_n 
\ee
The solution to this equation, with the initial condition $p_n(0)=\delta_{n1}$  is given by the distribution:
\begin{equation}\label{eq:bfklprob}
p_n(y)= \eto^{-\alpha y} \,(1-\eto^{-\alpha y})^{n-1}
\end{equation}
where n=1,2,\ldots , which is normalized to one,
\be
\sum_{n=1}^\infty p_n(y)=1
\ee
The mean multiplicity rises  exponentially with rapidity,
\be\label{eq:multbfkl}
\langle n\rangle=\sum_{n=1}^\infty n p_n(y)= \eto^{\alpha y}
\ee
and due to this dependence, the distribution (\ref{eq:bfklprob}) can be written as 
\be
p_n(y) =\frac{1}{\langle n\rangle} \left(1-\frac{1}{\langle n\rangle}\right)^{n-1}
\ee
With this result,   the entropy  reads:
\begin{align}\nonumber
S(y) &= -\sum_{n=1}^\infty p_n(y)\ln p_n(y)
\\
&=\nsr \ln \nsr -\left(\nsr-1\right) \ln\left(\nsr-1\right)
\end{align}
For  $y\to \infty$, when $\nsr \gg 1$,  the dipole probabilities become approximately uniform
\be\label{eq:pneq}
p_n\simeq \frac{1}{\nsr}
\ee
while  the entropy  is maximal and proportional to rapidity
\be\label{eq:bfklentropy}
S(y)\simeq  \ln \langle n\rangle = \alpha y
\ee
Thus, in contrast to the  model with recombination,  neither the mean multiplicity nor the entropy saturates. 

The exponential growth of the  mean multiplicity (\ref{eq:multbfkl})  is a characteristic feature of the  Balitsky-Fadin-Kuraev-Lipatov (BFKL) evolution in rapidity
\cite{Kuraev:1977fs,Balitsky:1978ic},  when  $\alpha=4\alpha_s\ln 2$ is the BFKL pomeron intercept. This evolution is known to violate the unitarity of the scattering matrix.  
Nevertheless, in the deep inelastic scattering of a dilute dipole projectile off a dense nuclear target, unitarity is restored through multiple dipole-nucleon scatterings \cite{Kovchegov:1999yj,Kovchegov:1999ua,Kovchegov:2012mbw}.

In  Ref.~\cite{Kharzeev:2017qzs}, the mean multiplicity in the model (\ref{eq:bfklmodel}) was identified with the mean number of gluons per unit of rapidity, $xg(x)$, found
from the BFKL equation,
\be
\langle n\rangle= xg(x) = x^{-\alpha}
\ee
where $x=\eto^{-y}\ll 1$.  
Thus, the entropy (\ref{eq:bfklentropy}) is given by 
\be\label{eq:59}
S(y)\approx \ln\, (xg(x))
\ee
This formula was further interpreted as entanglement  entropy. Since all partonic microstates have equal probabilities, see Eq.~(\ref{eq:pneq}),  the entanglement
entropy is maximal, and  the virtual probe in DIS at small $x$ sees the proton in a maximally entangled state.

As advocated in \cite{Kharzeev:2017qzs,Tu:2019ouv}, the physical picture underlying quantum entanglement is the division of the proton’s transverse area into  a subsystem $A$, of size $1/\sqrt{Q^2}$,  probed by a virtual electroweak boson, and the remainder of the proton  $B$. Averaging over
either of the two subsystems yields entanglement entropy, $S_A=S_B$. Therefore,  averaging over $B$ gives the partonic entropy
\be
S_A=S(y)
\ee
while averaging over $A$ gives the hadronic entropy
\be
S_B =S_{\text hadron} = -\sum_N P_N(y)\ln P_N(y)
\ee
where $P_N(y)$ is a probability distribution of $N$ hadrons (mostly charged pions) in the final state  at  rapidity $y$. Hence, one can identify the partonic and hadronic entropy
\be
\ln\, (xg(x)) \simeq S_{\text hadron}
\ee
This identification has been improved by including  quark distributions and  tested against experimental data in \cite{Tu:2019ouv,H1:2020zpd,Hentschinski:2022rsa,Hentschinski:2024gaa} with a positive result. 
An application of a generalized version of the model (\ref{eq:bfklmodel}) to the description of hadron multiplicities in the hadron-hadron scattering  data was presented in \cite{Kutak:2025tsx}.

A precise formulation of the entanglement  idea is still lacking, although the studies in this direction were carried out in \cite{Liu:2022qqf, Liu:2022hto,Liu:2022ohy} in two-dimensional QCD; see also \cite{Kharzeev:2026jkq}  for a recent, comprehensive review of the entanglement in high energy scattering.

\section{Conclusions}

We presented the concept of partonic entropy of the proton, defined in analogy to the Gibbs-Shannon entropy formula for a probability distribution, using the collinear parton distribution functions (PDFs). 
In this formulation, the probability distribution  $P(x)$ is given by the sum of the quark-singlet and gluon distributions multiplied by the Bjorken 
variable $x$, which is a positive quantity. With such a definition, the momentum sum rule (\ref{eq:mom}) is interpreted as the normalization condition for the probability distribution $P(x)$, which allows to define the non-negative partonic entropy  (\ref{eq:entropy}). 

We showed that within the DGLAP evolution scheme of the PDFs, 
the partonic entropy  is a monotonically increasing function of the evolution parameter $t=\ln(Q^2)$, where $Q^2$ is the QCD resolution (factorization) scale.  We also  showed that the partonic entropy  can be interpreted as a relative entropy of the density $P(x,t)$  with respect to  the uniform distribution on the interval $[0,1]$.

In order to better understand the origin of the monotonic growth of the partonic entropy, we presented a simplified model of DGLAP evolution in which  the  contribution from the region of $x\in [0,x_{\rm min}]$ is responsible for the unbounded growth of 
the partonic entropy. In this context, we highlighted the need to modify the small-$x$ behavior of the PDFs by incorporating parton saturation effects, which in turn leads to the saturation of the partonic entropy.

A precise formulation of parton saturation requires considering evolution equations with non-linear terms and is beyond  the scope on this work. Nevertheless, for completeness,   we presented a reaction-diffusion model of parton  saturation at small $x$, widely discussed in the literature, which leads to the saturation of  the partonic entropy. 

A truncated version of this model, without parton recombination, was also presented. Based on this simplified framework, it was conjectured 
that   the partonic entropy is essentially an entanglement  entropy.  In addition, in the limit of rapidity $y=\ln(1/x)\to \infty$,   the proton is probed in DIS   in a maximally entangled state, which  was  tested against the existing DIS data with a positive result.

\begin{acknowledgments}
We thank Krzysztof Kutak, Stanisław Mrówczyński and Anna Staśto for useful discussions. 
This work was supported by the  National Science Center, Poland,  Grants Nos. 2019/33/B/ST2/02588 and 2024/53/B/ST2/00968.
\end{acknowledgments}

\bibliographystyle{apsrev4-2}
\bibliography{mybib}

\end{document}